# Millimeter-scale focal length tuning with MEMS-integrated meta-optics employing high-throughput fabrication

Zheyi Han[1,4], Shane Colburn[1], Arka Majumdar[1,2], and Karl F. Böhringer[1,3,4,*]

[1]Department of Electrical and Computer Engineering, University of Washington, Seattle, WA 98195, USA
[2]Department of Physics, University of Washington, Seattle, WA 98195, USA
[3]Department of Bioengineering, University of Washington, Seattle, WA 98195, USA
[4]Institute for Nano-engineered Systems, University of Washington, Seattle, WA 98195, USA

*karlb@uw.edu

**Abstract:** Miniature varifocal lenses are crucial for many applications requiring compact optical systems. Here, utilizing electro-mechanically actuated 0.5-mm aperture infrared Alvarez meta-optics, we demonstrate 3.1 mm (200 diopters) focal length tuning with an actuation voltage below 40 V. This constitutes the largest focal length tuning in any low-power electro-mechanically actuated meta-optic, enabled by the high energy density in comb-drive actuators producing large displacements at relatively low voltage. The demonstrated device is produced by a novel nanofabrication process that accommodates meta-optics with a larger aperture and has improved alignment between meta-optics via flip-chip bonding. The whole fabrication process is CMOS compatible and amenable to high-throughput manufacturing.

## 1. Introduction

Miniature focus-tunable lenses are essential for many modern applications involving compact optical systems, such as augmented and virtual reality, microscopy, spectroscopy, sensing, and photonic integrated circuits. In many devices and instruments, the size and weight of the optics are highly constrained, while at the same time, low energy consumption and high controllability are often preferred attributes.

Among the many mechanisms developed to modulate the focusing power of lenses, a unique approach is based on an Alvarez lens[1,2]. A typical Alvarez lens consists of a pair of refractive optical elements with complementary cubic surface profiles. Focusing power is modulated by laterally displacing two optical elements perpendicular to the optical axis. The resultant focal length change is inversely proportional to the lateral displacement, and the focal length can be modulated in both directions. With the emergence and development of microelectromechanical systems (MEMS) for precision displacement, researchers have managed to assemble a pair of refractive Alvarez optics into a MEMS actuator to produce the required lateral displacement for focal tuning[3,4]; however, the refractive optics require complex, three-dimensional geometries incompatible with CMOS fabrication and hence are obstacles for ultimate miniaturization and direct integration in electronic systems.

Meta-optics are two-dimensional, quasi-periodic arrays of subwavelength scatterers designed to arbitrarily control electromagnetic waves' phase, amplitude, and polarization[5]. They present great potential to miniaturize bulky freeform optics with significantly reduced volume. Meta-optics can utilize well-established semiconductor microfabrication technologies for inexpensive, high-volume production and ready integration with electronic control. Their ability to produce the novel functionalities of freeform optics on a planar platform[6] has motivated the designs of meta-optical elements with a wide range of functionalities such as lenses[7-13], reflectors[14-16], vortex beam generators[17-20], holographic masks[21-23], blazed gratings[24,25] and polarization optics[26,27].

Recent advances in the design and fabrication of meta-optics have facilitated the active control and manipulation of light with tunable meta-optics. The application of external stimuli, including electrical[28], mechanical[29,30], thermal[31], optical[32], and magnetic[33] control, can dynamically alter the structure and characteristics of the meta-optics and modulate their interaction with electromagnetic waves. Owing to their exceptional capabilities in precise actuation and sensing within a confined space, there has been a rising interest in incorporating MEMS techniques to induce structural reconfiguration of the meta-optics and tune their optical properties in real-time. Such MEMS-actuated reconfigurations can either change the geometry and pitch of the scatterers in meta-optics by mechanical deformation[34,35] or adjust the relative positions of the whole meta-optic in a composite unit to modulate the overall optical properties[36-38]. Despite significant progress, the extent of the focal length modulation in a low-power MEMS-tunable metalens remains small (< 100 µm) when applying low to moderate tuning voltages. The main challenge to increase the tuning range while maintaining sufficient numerical aperture comes from the need to mechanically displace a large aperture meta-optic by a substantial amount, which makes the design and fabrication of MEMS structures difficult.

Here, we demonstrate a MEMS-tunable, 0.5-mm aperture infrared (IR) meta-optical lens exploiting the Alvarez principle. The reported focal length tuning range is 3.1 mm, approximately 40 times more than that of previous works[39,40]. Aperture size and tuning range increases are facilitated by improved mechanical structures, fabrication, and a novel bonding process. We demonstrate how the MEMS Alvarez lens's mechanical properties, actuation range, natural frequency, and tuning sensitivity can be readily modified by adjusting the comb-drive design parameters with minimal to no change in the device footprint. Additionally, the comb-drive actuators allow changing aperture size easily without drastically altering the actuator's operating mechanism or footprint. Such conveniences provide significant design flexibility either to accommodate particular application requirements or to improve focal tuning efficiency. The entire fabrication process is CMOS compatible using common semiconductor materials without involving electron-beam lithography, suitable for potential high-volume, low-cost mass production.

## 2. Design

The design of MEMS-integrated Alvarez meta-optics involves two main parts. First, we map the two complementary cubic surfaces onto the quasi-periodic arrays of flat meta-optics. Second, we design electrostatic actuators to induce the required lateral displacement to modulate the optical power of the assembled Alvarez lens.

### 2.1 Alvarez meta-optics

We have designed a pair of square Alvarez meta-optics with an aperture size of 0.5 mm. They host complementary cubic surface profiles for a designed operating wavelength of 1550 nm. The regular and inverse surfaces have the phase distributions

$$\varphi_{reg}(x,y) = -\varphi_{inv}(x,y) = A\left(\frac{1}{3}x^3 + xy^2\right). \tag{1}$$

where $(x, y)$ represents the in-plane coordinates. The constant $A$ denotes the cubic phase strength determining the rate of phase variation with the units of inverse cubic length[6,41].

As illustrated in Figure 1(a), when overlapped with a nonzero center-to-center offset, the two cubic surfaces in conjunction impart a quadratic total phase profile on an incident wavefront, with the form

$$\varphi_{Alvarez}(x,y) = \varphi_{reg}(x+d,y) + \varphi_{inv}(x-d,y) = 2Ad(x^2+y^2) + \frac{2}{3}Ad^3, \tag{2}$$

whose focal length is tunable with laterally displacing the two surfaces in the opposite direction by an offset of $d$ each (giving a total center-to-center offset of $2d$). With the constant $d^3$ term being inconsequential, the quadratic term of the total Alvarez phase profile can be related to that of a standard quadratic lens, giving the reciprocal relation between the tunable focal length *f* and symmetric lateral displacement *d* as

$$f(d) = \frac{\pi}{2\lambda A d}. \quad (3)$$

We chose cylindrical nanoposts fabricated using silicon nitride on a silicon substrate as the scattering elements, as shown in Figure 1(b). Via rigorous coupled-wave analysis (RCWA), the transmission coefficients and phase shifts of the nanoposts for a fixed lattice constant of 1.3 µm and a post height of 2 µm are simulated as a function of the grating duty cycle, defined as the ratio between the post diameter and pitch. The calculated phase spans from 0 to $2\pi$ while maintaining near-unity transmission amplitude, as shown in Figure 1(c). The phase profile is then quantized into six linear steps of duty cycle, giving the corresponding six cylindrical post diameters achievable with high-throughput stepper lithography [41] to construct the cubic phase profiles on the Alvarez meta-optics. We introduce an initial 40 µm center-to-center offset between two meta-optics. The focal length is then further tuned by introducing an additional displacement to this built-in offset via MEMS actuation.

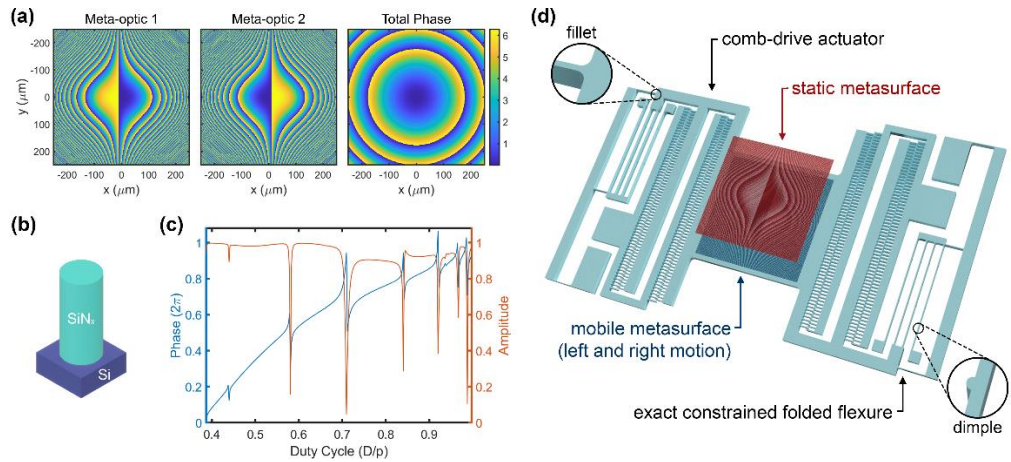

**Figure 1.** Design of the MEMS-integrated Alvarez meta-optic. (a) The designed cubic phase profiles for the two complementary Alvarez meta-optics and the total quadratic phase profile when they overlay. Colorbar indicates a $2\pi$ phase span in radian. (b) Schematic of the scatterer made of a cylindrical silicon nitride nanopost on a silicon substrate. (c) The simulated transmission coefficients for the nanoposts as a function of the duty cycle. (d) Schematic of Alvarez meta-optics integrated with an electrostatic MEMS actuator.

## 2.2 MEMS electrostatic tuning platform

The conventional configuration of the Alvarez lens displaces two complementary cubic optics symmetrically in opposite directions to modulate the center-to-center offset for focal tuning. Here, to ensure fabrication robustness, we chose to alter the conventional mechanism to actively displace only one meta-optic relative to its static complement, producing a net change in the center-to-center offset between the two cubic surfaces.

Comb-drive actuators exemplified in **Figure 1**(d) have high voltage-force conversion efficiencies realized by increased edge coupling length with the interdigitated fingers[42]. Hence, we employ them to electrostatically control the lateral displacement of the mobile meta-optic. The meta-optic sits on a central platform with comb drives on opposite sides to control the displacement in both positive and negative directions along the actuation axis. The symmetric design enhances the mechanical stability and doubles the range of actuation without doubling the maximum voltage required. The magnitude of electrostatic force $F_{el}$ driving the lateral motion is quadratically dependent on the applied voltage as

$$F_{el}(V) = \frac{N\varepsilon h V^2}{d_{sep}}, \qquad (4)$$

where $N$ is the number of finger pairs, $\varepsilon$ is the electric permittivity of the dielectric medium (air in this case), $h$ is the finger height, and $d_{sep}$ is the separation between the neighboring fingers.

To generate the restoring mechanical force in the actuator, we design a modified version of the exact constrained folded flexures[43] for their sideways stiffness stability at relatively large displacement. As shown in the circular insets of **Figure 1**(d), we introduce fillets (rounding of corners) at mechanical hot spots to reduce, redistribute and partially relieve stress created during actuation and prolong device lifetime especially for many-cycle operations. We also introduce horizontal dimples (small protrusions) sparsely placed along the backbones and long springs in the folded flexures, to prevent structural sticking if accidentally running into contact during large-displacement operation.

By Hooke's law, the displacement is linearly proportional to the spring force in the flexure with a spring constant $k_{sp}$. Hence at the equilibrium between the electrostatic driving force and restoring spring force, the actuated lateral displacement $\Delta d$ is quadratically dependent on the applied voltage following the relation

$$\Delta d = \frac{F_{el}(V)}{k_{sp}} = \frac{N\varepsilon h V^2}{k_{sp} d_{sep}}. \qquad (5)$$

Such direct dependence of the displacement on the driving voltage provides for high reproducibility and controllability in any calibrated system. Because of the electrostatic actuation mechanism, no current flows in the device when applied with direct-current (DC) voltages, enabling very low-power consumption at no more than several nanowatts.

## 3. Methods

Subsequently, we fabricate Alvarez meta-optics with two different actuator designs and characterize their electrostatic and optical performances.

### 3.1 Device fabrication

We fabricate two complementary Alvarez meta-optics on two separate substrates before aligning and bonding them to form the final Alvarez lens. **Figure 2**(a) summarizes the fabrication process flow. Given that we are integrating a significantly larger meta-optic on the MEMS platform and using modestly higher actuation voltages to induce larger displacements, to alleviate any out-of-plane sagging offset of the release optics due to imbalanced fringing fields, a thicker silicon layer of 11 μm is employed for the actuator. Similarly, to assist a more efficient and stable release process for the suspended flexures, we have carefully designed the release hole patterns to create a more even exposure for release reactions while still ensuring strong mechanical stability within the flexures. Finally, we employ a flip-chip bonding process to reduce the vertical separation down to 10 μm and lateral misalignment to less than 3 um between the two meta-optics.

We fabricate the mobile meta-optic on a silicon-on-insulator (SOI) wafer with the electrostatic actuators. A 2 μm thick silicon nitride layer is first deposited with plasma-enhanced chemical vapor deposition (PECVD) on the SOI wafer. We employ i-line stepper lithography to pattern the meta-optic with high throughput. The meta-optic pattern is then transferred to the silicon nitride layer via fluorine-based inductively coupled plasma (ICP-F) etching utilizing an aluminum hard mask. A metal stack consisting of 10 nm thick chromium and 150 nm thick gold is evaporated and lifted off to form the contact lines and probing pads defined by direct-write lithography. The 11 μm thick silicon device layer is then patterned with the actuator design using direct-write lithography and etched with deep reactive-ion etching (DRIE). We use direct-write lithography for quick prototyping of various actuator designs. This step can be easily replaced by more efficient flash exposure processes such as stepper lithography for cost efficiency in potential high-throughput production. After the actuator fabrication, a backside DRIE step is performed to remove the 400 µm handle silicon under the central meta-optic plate to aid the later release process. The wafer is diced into individual chips containing multiple devices before being etched in vapor hydrogen fluoride (vHF) to remove the exposed buried oxide (BOX) and release the actuator with the mobile meta-optic.

We fabricate the static meta-optic on a 525 µm thick double-side polished silicon wafer. Similarly, a 2 µm layer of PECVD silicon nitride is deposited, and the static meta-optic is patterned with stepper lithography followed by ICP-F etching into the thin film. The wafer is then diced into individual chips, each containing multiple meta-optics corresponding to their mobile counterparts.

The mobile meta-optic on the MEMS actuator and the static meta-optic are then assembled to create the complete Alvarez unit. For an Alvarez lens, it is crucial to have a small gap between two optics [6]. To ensure that, the chip with mobile meta-optics and the one with static meta-optics are aligned and bonded on a flip-chip bonder, using pieces of an anisotropic conductive film (ACF) both as the spacer and the adhesive layer to create a significantly smaller axial gap with high controllability. This commercial ACF contains polymer particles in a thermoplastic adhesive matrix. The spherical particles have radii around 10 µm. After applying heat and pressure, the adhesive spacer flows to a thickness measured to be approximately 10 µm and cross-links, securing the two chips in an aligned stack with a minimal gap sufficient to prevent meta-optic scratching during operation.

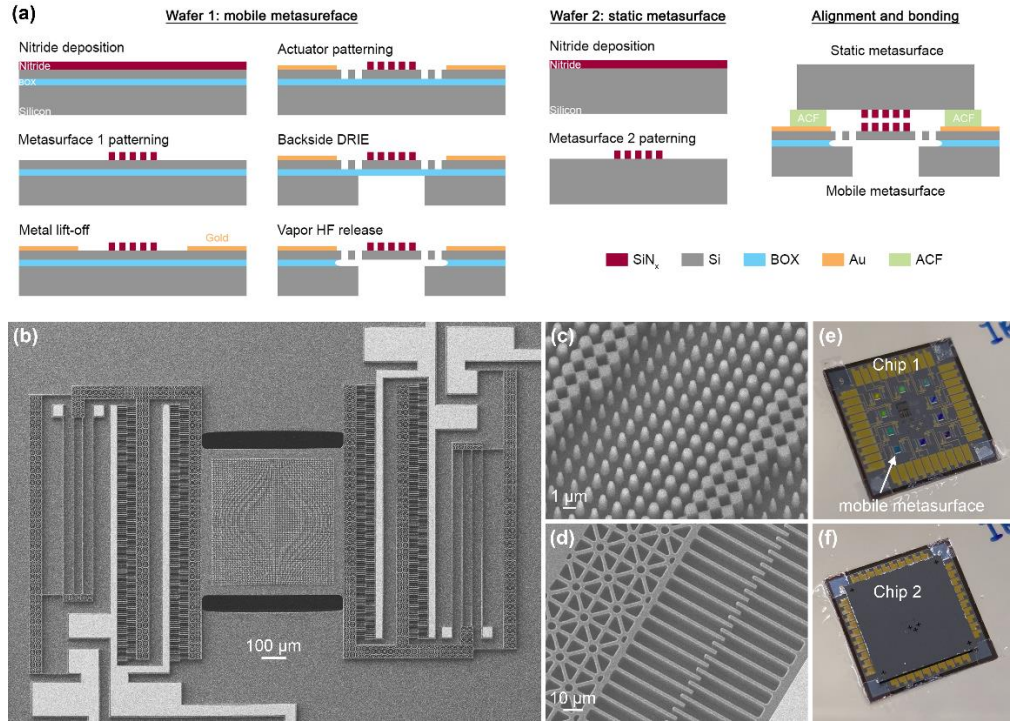

**Figure 2.** Device fabrication. (a) Summarized fabrication process flow. (b) Scanning electron microscopy (SEM) image of an Alvarez meta-optic integrated with a MEMS tuning platform. (c) Close-up view of the silicon nitride nanoposts sitting on the central silicon platform. (d) Comb-drive details showing the part of the mobile flexure backbone and interdigitated finger array. (e) Fabricated chip carrying mobile Alvarez meta-optics on electrostatic actuators. (f) Final assembled stack with another chip carrying static Alvarez meta-optics aligned and bonded on top.

**Figure 2**(b) shows the fabricated mobile meta-optic on the central silicon platform of a comb-drive actuator before bonding with its static counterpart. **Figure 2**(c) zooms into a portion of the meta-optic consisting of silicon nitride nanoposts with varying diameters arranged in a two-dimensional quasi-periodic array to construct one Alvarez meta-optic. **Figure 2**(d) captures the interdigitated fingers that generate the electrostatic force driving the lateral displacement upon voltage application and part of the release holes designed to assist the vHF process in releasing the mobile half combe-drive for actuation. **Figure 2**(e) captures an SOI chip carrying 12 mobile meta-optics on electrostatic actuators, each having an overall device footprint of 2 mm × 1.2 mm. **Figure 2**(f) shows the final Alvarez stack when the second chip with 12 complementary stationary meta-optics is aligned and bonded face-down with its mobile counterpart. The exposed contact pads along the chip edges are used for electrical probing during the focal tuning tests. The measurements at the chip-to-chip alignment marks confirm that the translational misalignment does not exceed 3 µm in all directions, and the rotational misalignment is less than 0.03° between the two meta-optics. We have also fabricated another actuator design with fewer comb fingers of larger width and shorter springs for higher stiffness and hence higher resonant frequency. The corresponding images and detailed results are provided in the Supplement. Table 1 summarizes the dimensions and performances of the two designs.

|  |  | Design 1 | Design 2 |
|---|---|---|---|
| Comb Fingers | Width (µm) | 2 | 3 |
|  | Length (µm) | 40 | 40 |
|  | Height (µm) | 11 | 11 |
|  | Gap (µm) | 2 | 2 |
| Folded Springs | Width (µm) | 4 | 4 |
|  | Length (µm) | 700 | 500 |
|  | Height (µm) | 11 | 11 |
|  | Lateral Stiffness (N/m) | 1.15 | 2.29 |
|  | Natural Frequency (Hz) | 1300 | 1800 |
| Focal Tuning | Voltage (V) | −10 to +40 | −15 to +55 |
|  | Displacement (µm) | −1.2 to +18.1 | −0.5 to +13.1 |
|  | Focal Length (mm) | 5.8 to 2.7 | 6.0 to 3.0 |

**Table 1.** Design parameters and performances of the MEMS Alvarez meta-optic lens. Device 2 images and results are presented in the Supplement.

## 4. Experimental results

We test and evaluate the electrostatic performance of the MEMS actuator for its capability to create controlled uniaxial displacement required for Alvarez tuning. We then monitor and analyze the focal tuning behaviors of the lens when different lateral displacements are introduced at different actuation voltages. **Figure 3**(a) illustrates the general experimental setup for all the measurements. The chip stack containing Alvarez meta-optic lenses is affixed using transparent ultraviolet (UV) resin to a microscope slide, which is clamped vertically on a 3-axis micro-manipulator stage, perpendicular to the optical axis. We probe the Alvarez lens with a DC voltage to laterally actuate the mobile meta-optic relative to the static one. A 1550 nm superluminescent diode (SLD) source illuminates the Alvarez lens through a collimator. The incident light interacts with the meta-optic stack and converges to a focal spot along the optical axis on the other side. An IR microscope setup on a translation stage is used to measure the in-plane intensity profiles.

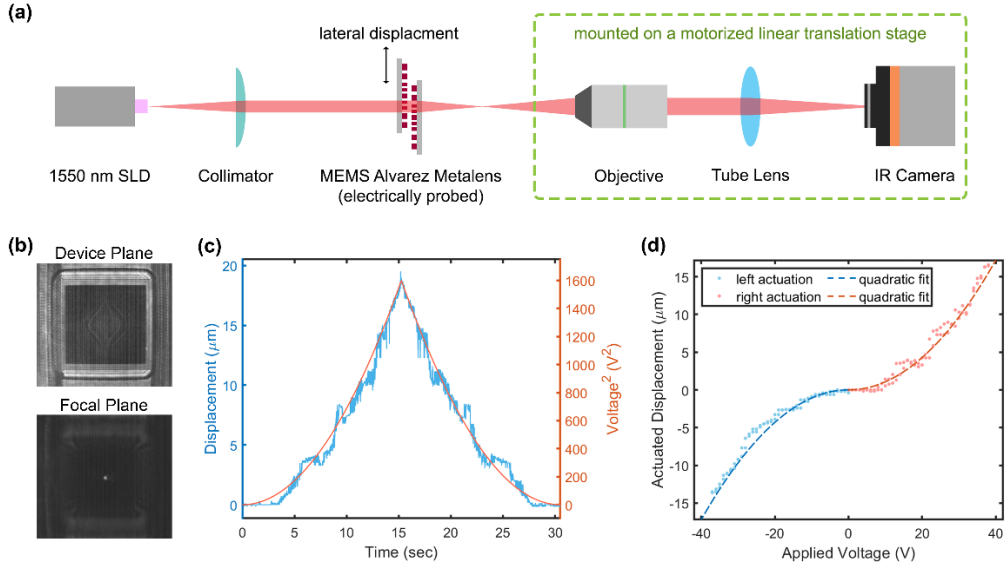

**Figure 3.** Experimental characterization of the Alvarez lens. (a) Experimental setup for electrostatic actuation and focal profile acquisition. (b) Exemplary device and focal planes of the Alvarez meta-optic lens actuated at 40 V towards the right. (c) Measured actuated displacement and actuating voltage of the MEMS platform follow the same trend closely, showing negligible hysteresis. (d) Actuated displacement for both directions follows the comb-drive quadratic characteristics closely.

## 4.1 Electrostatic tuning

Since the electrostatic actuators in our MEMS Alvarez lens consist of comb-drives on both left and right of the central plate, enabling actuation in both directions, we use sign conventions to differentiate the two sides. The positive voltage indicates the stimulus applied on the right comb drives. The corresponding positive displacement pulls the mobile meta-optic towards the right, increasing the center-to-center offset between the Alvarez meta-optic pair. In contrast, negative voltage and displacement values are attributed to the left, reducing the center-to-center offset. To evaluate the performance of electrostatic actuation, we apply a continuous linear voltage ramp from 0 V to 40 V and then back down to 0 V on both sides of the actuator respectively and monitor the lateral displacement at the focal plane via video recording. **Figure 3**(b) shows the representative screenshots displaying the device plane and the corresponding focal plane when the mobile meta-optic is pulled electrostatically towards the right at 40 V. We measure the lateral displacement of the mobile meta-optic at any instance in the video by analyzing the comb-drive motion using edge detection.

**Figure 3**(c) plots the displacement of the mobile meta-optic and the corresponding actuation voltage on the right side during the ramping process. The displacement increases from 0 µm to 18 µm before decreases as the applied voltage rises from 0 V to 40 V and then goes back down. The symmetric local kinks and noise in the displacement curve result from the camera resolution limit and the image intensity nonuniformity due to residual camera artifacts interfering with the edge detection process. The actuated displacement follows closely with the square of the actuation voltage during both ramp-up and ramp-down with negligible hysteresis, granting high controllability and reproducibility of the tuning process using such comb-drive actuators. **Figure 3**(d) plots the displacement versus actuation voltage for actuation on both left and right sides of the same MEMS Alvarez lens. All the data points lie close to the characteristic quadratic displacement-voltage dependency presented in Eq. (5), with some minor deviations

between the two sides due to fabrication imperfections. The quadratic fitting gives a measured spring constant of 1.15 N/m, close to the analytical estimate of 1.07 N/m. The analytical estimates for the spring constants are based on the folded flexures[44], which effectively determine the compliance in the actuator. The minor discrepancies between the measured and analytical values can be attributed to fabrication imperfections and the addition of side levers for enhanced sideways stability[43]. Compared to the analytical stiffness of 32,700 N/m in the perpendicular direction, the stiffness ratio between the perpendicular and the actuating directions is as high as 28,500. The second design with shorter springs has a higher measured lateral spring constant of 2.29 N/m (close to its analytical estimate of 2.93 N/m) and an analytical perpendicular stiffness of 45,800 N/m, giving a stiffness ratio of 20,000. Such high stiffness ratios between the perpendicular and lateral directions ensure high robustness during actuation, providing controlled uniaxial displacement essential for tuning Alvarez meta-optics. We note that the stiffness along the perpendicular direction could not be measured in the experiments due to minimal displacement along that direction, which confirms the high-stiffness-ratio design for highly controllable uniaxial actuation.

Both actuator designs have a total mass of mobile structures around $m = 18$ μg. With the measured spring constants of $k_{sp}^{(1)} = 1.15$ N/m and $k_{sp}^{(2)} = 2.29$ N/m, we can calculate the natural frequency of the actuators as

$$f_0 = \frac{1}{2\pi}\sqrt{\frac{k_{sp}}{m}}, \tag{6}$$

giving the corresponding natural frequencies as $f_0^{(1)} = 1.3$ kHz and $f_0^{(2)} = 1.8$ kHz for the two designs, showing potential for kHz operation. The comparison shows that although the second actuator design has a stiffer structure and requires slightly higher actuation voltage, its resonance will occur at a higher frequency, making the device more robust over a wider operating frequency range. Hence, the actuator designs can be flexibly modified within the same footprint to boost specific attributes to fulfill different application requirements.

Section 2 of the Supplement shows the calculations and plots for device power consumption. For all the operations under DC voltages, the current values detected are in nano-amperes, which are probably due to minor leakage current via the substrate. The resultant powers are no more than 10 nW, realizing very low power consumption with DC operation. We estimate the power consumption at higher operating frequencies by calculating the capacitive energy consumed per switching (i.e., per period) based on the measured properties and design dimensions. With a natural frequency of 1.3 kHz, the MEMS Alvarez lens presented in this paper is estimated to have a maximum switching power consumption below 1 μW, realizing low power consumption for kHz tuning.

### 4.2 Alvarez focal tuning

To tune the focal length, we apply a DC voltage across the comb drives on one side of the mobile meta-optic, displacing it laterally to modulate the center-to-center offset between the two Alvarez meta-optics. The lateral displacement of the mobile meta-optic is found by using edge detection in the device plane. Then the linear translation stage moves the IR microscope away from the device to find the region where the transmitted light focuses. The microscope scans along the optical axis with increments of 5 μm to capture the intensity profiles within a total of 0.8 mm span around the main intensity lobe to search for the focal plane. We apply an actuation voltage up to 40 V to pull the mobile meta-optic towards the right – indicated by positive voltage and displacement values – before reaching the regime of pull-in instability. Then we apply an actuation voltage up to 10 V to pull it towards the left – indicated by negative voltage and displacement values. As the offset decreases, the focal length increases, eventually making the focal spot large and dim.

**Figure 4**(a) displays the changes in focal profile and location along the optical axis (z-axis) at varying actuation voltage and actuated displacement, which modulate the center-to-center offset between the two Alvarez meta-optics. We define the focal length as the distance from the device (located at $z = 0$ mm plane) to the *x-y* plane, where we have the highest intensity spot. The square images at the right side of each row show the corresponding intensity distributions at the focal planes. We note that a higher actuation voltage induces a larger lateral displacement, which translates to a larger center-to-center offset between the two meta-optics, resulting in shorter focal length and higher effective numerical aperture. Hence, we obtain a sharper focal spot with higher actuation voltage, as shown in the inset of Fig. 4(a). We also observe some aberrations in the focal spot shape, primarily coming from the slight misalignment between two meta-optics.

**Figure 4**(b) plots the actuated displacement and the resultant focal length as functions of the applied voltage. As the applied voltage increases from –10 V to +40 V (sign indicating actuation directions only), the induced lateral displacement in the mobile meta-optic increases from –1.2 µm to +18.1 µm, and their absolute values show a quadratic dependence on the voltages, agreeing with the theoretical behavior of comb-drive actuators. The lateral displacements add onto the built-in 40 µm center-to-center offset between the Alvarez meta-optics and modulate the focal length from 5.8 mm to 2.7 mm, producing a total focal tuning range of 3.1 mm and a change in optical power of 200 diopters. The dashed lines plot the fitting to the corresponding characteristic behaviors, with inflection occurring at 0 V. The displacement-voltage data closely follows the quadratic behavior described in Eq. (5), and the focal-voltage data follows the relation predicted by the combination of Eq. (3) and Eq. (5). We note that the focal length shift at around 20 V is slightly off, which we attribute to experimental imperfections. Similarly, the second type of actuator (numerical results in Table 1 and plots in the Supplement) produces a focal length tuning of 3.0 mm and an optical power change of 166 diopters, realized by lateral displacements ranging from –0.5 µm to +13.1 µm induced by voltages between –15 V and +55 V.

**Figure 4**(c) plots the focal length *f* versus the actuated displacement *Δd*, which follows the adapted theoretical Alvarez tuning behavior closely as

$$f(\Delta d) = \frac{\pi}{2\lambda A \left( d_0 + \Delta d \right)}, \tag{7}$$

where $d_0 = 40$ µm is the initial center-to-center offset introduced in the designs of Alvarez meta-optics and $\lambda = 1550$ nm is the operating wavelength. The good fit between experiment and theory confirms the successful adaptation of a conventional Alvarez system to a miniature on-chip system via MEMS-integrated meta-optics. The experimental data give an actual value for the cubic phase strength *A* as $(5.42 \pm 1.08) \times 10^{12}$ m$^{-3}$.

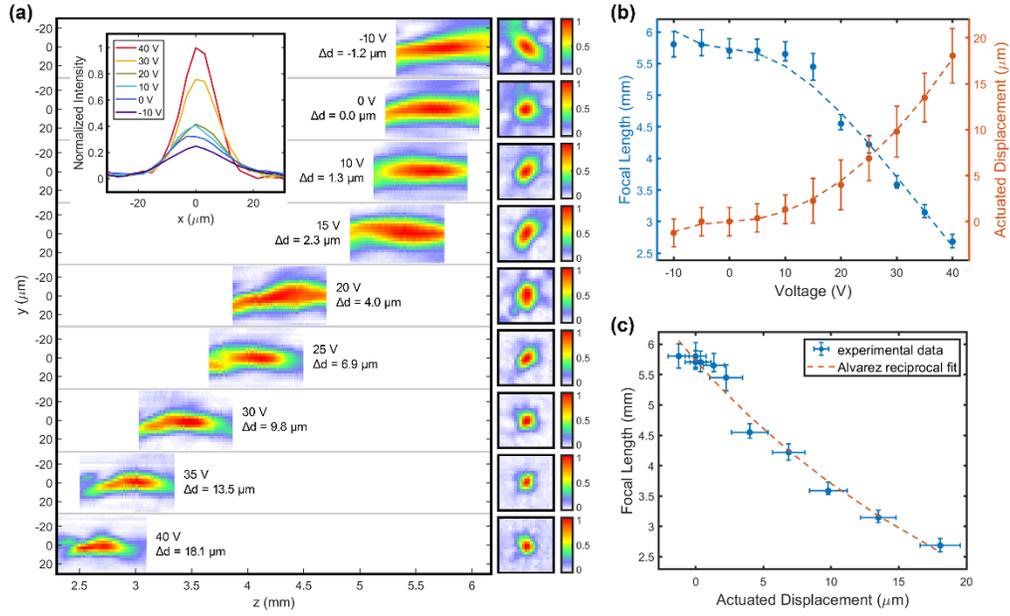

**Figure 4.** Focal tuning measurements of the MEMS Alvarez meta-optical lens. (a) Normalized focal profiles along the optical axis (z-axis) capturing the intensities at 5 µm increments across a 0.8 mm span around the main intensity lobes. Square plots on the right show the intensity distributions in the found focal planes. Inset: normalized intensities across the focal spots at various actuation voltages. (b) Actuated displacement and the corresponding focal length modulated by actuation voltage. Dashed lines show the fitting to the theoretical behaviors. (c) Tunable focal length as a function of actuated displacement closely follows the theoretical reciprocal Alvarez tuning behavior.

## 5. Discussion and conclusion

We demonstrated a miniature varifocal lens using MEMS-integrated meta-optics exploiting the concept of Alvarez lens. The efficient comb-drive actuation produces a maximum displacement range of 19 µm with input voltages below 40 V. The inverse-dependence of the focal length on the displacement in an Alvarez lens enables a focal tuning by 3.1 mm, more than an order of magnitude larger than the previous reports. Thanks to the electrostatic actuation, the consumed power is lower than 10 nW for DC operation and less than 1 µW for higher tuning frequencies into kHz. Additionally, the whole fabrication process is compatible with high-volume manufacturing, making such an integrated platform attractive for various applications requiring miniature tunable free-space optics.

The complementary cubic surface profiles in the Alvarez system require accurate alignment to achieve optimal performance. While translational misalignment along the actuation direction can be compensated by electrostatic displacement, misalignment along the perpendicular or rotational direction is more challenging to correct in the assembled stack. Though small, the translational misalignment (< 3 µm) and rotational misalignment (< 0.03°) in our device still might have contributed to the non-spherical foci at some voltages, where the cubic surface terms fail to cancel each other altogether when overlapping. An alternative chip and bonding process with better accuracy together with a meta-optic design upgrade with higher misalignment tolerance may provide better results. Like other meta-optics, our varifocal lens suffers from chromatic aberration, limiting the operation to a narrow bandwidth around the design wavelength. Possible improvement can involve further engineering of the meta-optics, such as the quartic meta-optics, which have been recently employed to simultaneously achieve achromatic operation and varifocal control at visible wavelengths[45].

**Data availability**

Data underlying the results presented in this paper are available from the corresponding author upon request.

## Funding

Tunoptix Inc. and DARPA (Contract no. 140D0420C0060)

## Acknowledgments

This work was supported by funding provided by Tunoptix Inc. and DARPA (Contract no. 140D0420C0060). Part of this work was conducted at the Washington Nanofabrication Facility / Molecular Analysis Facility, a National Nanotechnology Coordinated Infrastructure (NNCI) site at the University of Washington with partial support from the National Science Foundation via awards NNCI-2025489, NNCI-1542101 and ECCS-1337840.

## Author contributions

Z.H. designed, fabricated, and characterized the integrated miniature MEMS-actuated focal-tunable lens incorporating the Alvarez meta-optics designed by S.C. K.B. and A.M. supervised the MEMS integration and Alvarez meta-optics design respectively as the principal investigators.

## Competing interests

A.M., S.C. and K.B. are part of the company Tunoptix, which is commercializing related technology. Z.H. declares no potential conflict of interest.

**Legends**

**Figure 1.** Design of the MEMS-integrated Alvarez meta-optic. (a) The designed cubic phase profiles for the two complementary Alvarez meta-optics and the total quadratic phase profile when they overlay. Colorbar indicates a 2π phase span in radian. (b) Schematic of the scatterer made of a cylindrical silicon nitride nanopost on a silicon substrate. (c) The simulated transmission coefficients for the nanoposts as a function of the duty cycle. (d) Schematic of Alvarez meta-optics integrated with an electrostatic MEMS actuator.

**Figure 2.** Device fabrication. (a) Summarized fabrication process flow. (b) Scanning electron microscopy (SEM) image of an Alvarez meta-optic integrated with a MEMS tuning platform. (c) Close-up view of the silicon nitride nanoposts sitting on the central silicon platform. (d) Comb-drive details showing the part of the mobile flexure backbone and interdigitated finger array. (e) Fabricated chip carrying mobile Alvarez meta-optics on electrostatic actuators. (f) Final assembled stack with another chip carrying static Alvarez meta-optics aligned and bonded on top.

**Figure 3.** Experimental characterization of the Alvarez lens. (a) Experimental setup for electrostatic actuation and focal profile acquisition. (b) Exemplary device and focal planes of the Alvarez meta-optic lens actuated at 40 V towards the right. (c) Measured actuated displacement and actuating voltage of the MEMS platform follow the same trend closely, showing negligible hysteresis. (d) Actuated displacement for both directions follows the comb-drive quadratic characteristics closely.

**Figure 4.** Focal tuning measurements of the MEMS Alvarez meta-optical lens. (a) Normalized focal profiles along the optical axis (z-axis) capturing the intensities at 5 µm increments across a 0.8 mm span around the main intensity lobes. Square plots on the right show the intensity distributions in the found focal planes. Inset: normalized intensities across the focal spots at various actuation voltages. (b) Actuated displacement and the corresponding focal length modulated by actuation voltage. Dashed lines show the fitting to the theoretical behaviors. (c) Tunable focal length as a function of actuated displacement closely follows the theoretical reciprocal Alvarez tuning behavior.

**Table 1.** Design parameters and performances of the MEMS Alvarez meta-optic lens. Device 2 images and results are presented in the Supplement.